\documentclass[aps,prl,showpacs,twocolumn,superscriptaddress,nofootinbib,floatfix,mamsmath,10pt]{revtex4}
\usepackage{epsfig,epsf,graphics,psfrag}
\usepackage{times}
\usepackage{float}
\usepackage{color}
\usepackage{amsfonts,amssymb,stmaryrd,latexsym,amsmath}

\newcommand{\be}{\begin{equation}}
\newcommand{\ee}{\end{equation}}
\newcommand{\ba}{\begin{eqnarray}}
\newcommand{\ea}{\end{eqnarray}}

\begin{document}
%\preprint{}
\title{Reduction of the proton radius discrepancy by 3 {\boldmath$\sigma$}}

\author{I.~T.~Lorenz}
\email{lorenzi@hiskp.uni-bonn.de}
\affiliation{Helmholtz-Institut f\"ur Strahlen- und
             Kernphysik and Bethe Center for Theoretical Physics, \\
             Universit\"at Bonn,  D--53115 Bonn, Germany}

\author{Ulf-G.~Mei{\ss}ner}
\email{meissner@hiskp.uni-bonn.de}
\affiliation{Helmholtz-Institut f\"ur Strahlen- und
             Kernphysik and Bethe Center for Theoretical Physics, \\
             Universit\"at Bonn,  D--53115 Bonn, Germany}
\affiliation{Institute for Advanced Simulation, Institut f\"{u}r Kernphysik
             and J\"ulich Center for Hadron Physics, \\
             Forschungszentrum J\"{u}lich, D--52425 J\"{u}lich, Germany}

\begin{abstract}
\noindent We show that in previous analyses of electron-proton scattering, the
uncertainties in the statistical procedure to extract the
proton charge radius  are underestimated.  Using 
a fit function based on a conformal mapping, we can describe  the
scattering data with high precision and extract a radius value 
in agreement with the one obtained from muonic hydrogen.

\end{abstract}

%\pacs{13.40.Gp, 14.20.Dh, 11.55.Fv}

\maketitle

{\bf 1.}
Two principally different methods are commonly used to determine the proton 
charge radius $r_E^P$. On the one hand, it enters 
the QED calculations of atomic energy splittings (electronic and muonic \cite{Pohl:2010zza}
hydrogen) and can thus be obtained from measurements of these. 
On the other hand, $r_E^P$ can be obtained from elastic electron-proton
scattering. The corresponding cross sections can be parameterized in terms of 
the electric and magnetic Sachs form factors $G_E(Q^2)$ and $G_M(Q^2)$,
respectively, that depend on the invariant momentum transfer squared $Q^2 =
-t$. Positive $Q^2$-values refer to the scattering process, negative to
annihilation/creation. The reduced cross section, here in the one-photon
approximation,
describes the deviation from the scattering off a point-like particle: 
\begin{equation}\label{eq:xs_ros}
\frac{d\sigma}{d\Omega}\biggr|_{\rm red} = \frac{\tau}{\epsilon (1+\tau)}
\left[G_{M}^{2}(Q^{2}) + \frac{\epsilon}{\tau} G_{E}^{2}(Q^{2})\right]\, ,
\end{equation}
where $\epsilon = [1+2(1+\tau)\tan^{2} (\theta/2)]^{-1}$ is the virtual 
photon polarization, $\theta$ is the electron scattering angle in the 
laboratory frame and $\tau = -t/4m_N^2$, with $m_N$ the nucleon mass.\newline
Both methods refer to the same quantity, the slope of the proton form 
factor at the origin:
\begin{equation}\label{eq:rad}
r_{E,M}^p = \left(\left.\frac{-6}{G_{E,M}(0)}\frac{dG_{E,M}(Q^2)}{dQ^2}
\right|_{Q^2 = 0}\right)^{1/2}~.
\end{equation}
The form factor obtained from the cross sections has to be extrapolated from 
the data at lowest momentum transfer to the origin.
The most precise electron-proton scattering data from Ref.~\cite{Bernauer:2010wm}
analyzed using spline and polynomial fit functions lead to a proton charge 
radius that differs by $\sim 7\,\sigma$ from the muonic hydrogen
radius of Ref.~\cite{Pohl:2010zza}, when averaged with measurements in electronic 
hydrogen \cite{Beyer13}.
\newline 
The purpose of this letter is to illustrate that such extrapolations 
lack precision in purely statistical analyses with arbi-
trary fit functions. For example, the fit functions quoted in the final results of 
Ref.~\cite{Bernauer:2010wm} are polynomials and splines. 
In this letter, we construct a simple function, that
describes the data {\em equally well} and 
corresponds to a small radius $r_E^P$ in agreement with the one obtained from 
muonic hydrogen spectroscopy.
This function is based on a conformal mapping and thus obeys the analytic structure 
of the form factors. The following function maps the cut in the $t$-plane onto
the unit circle in a new variable $z$:
\begin{equation}
 z(t,t_{\rm cut})=\frac{\sqrt{t_{\rm cut}-t}-\sqrt{t_{\rm cut}}}{\sqrt{t_{\rm
       cut}-t}+\sqrt{t_{\rm cut}}}~,
\end{equation}
where $t_{\rm cut} = 4M_{\pi}^2$ is the lowest singularity of the form factors 
with $M_{\pi}$ the charged pion mass. The Sachs form factors can then be expanded in 
the new variable $z$:
\begin{equation}
 G_{E/M}(z(t))=\sum_{k=0}^{k_{\rm max}}a_kz(t)^k~.
\label{expa}
\end{equation}
Here, the form factors are normalized to the charge and anomalous magnetic 
moment of the proton, respectively.\newline
Conformal mapping techniques are a standard tool in hadron physics. So far, 
they have not been applied to the electron-proton scattering data by the 
A1-collaboration, the data for this process with the highest quoted
precision.  A previous elaborate analysis of world form factor data in a similar approach 
was carried out by Hill and Paz \cite{Hill:2010yb}. In contrast to their analysis,
we do not constrain the parameters $a_k$ any further to have a most flexible fit
function, which is needed for the statistical reasoning here. Moreover, the results by Hill and Paz refer to older form factor data, that are extracted from cross sections mainly via the Rosenbluth method. We avoid the systematical uncertainties related to this procedure by directly fitting the cross sections. Also, the results by Hill and Paz show a strong ambiguity due to the included fit range. As we have shown before \cite{Lorenz12}, this can be avoided in a full dispersion relation approach, since this makes use of the complete available information on the spectral function. Loose constraints on the coefficients in a $z$-expansion neglect the mass-related information on the spectral function.
\newline
We emphasize the mainly illustrative purposes of this work. This means that 
the significance of these fits lies in the comparison to the data analysis by
the A1-collaboration \cite{Bernauer:2010wm}. To allow for a direct comparison 
to that work, we use exactly the same data without further radiative
corrections and with fixed normalization parameters (see the next section for 
details). Physically, the main advantage of the
function used here, compared to the polynomials and splines used by the
experimenters, is the correct inclusion of the lowest singularity of the form
factors. Besides the basic analytic structure of the form factors, the
imaginary part of the form factors can be constrained further due to
unitarity, see e.g. \cite{Lorenz12}, but this goes beyond the aim of this work.  
However, arbitrary coefficients correspond to an unconstrained spectral function. 
Therefore, the fits shown here have to be treated on the same footing as polynomial or spline fits.

\medskip
\noindent{\bf 2.}
%----------------------------------------------------------------------------------
\begin{figure}[t]
\centering
\includegraphics[width=0.48\textwidth]{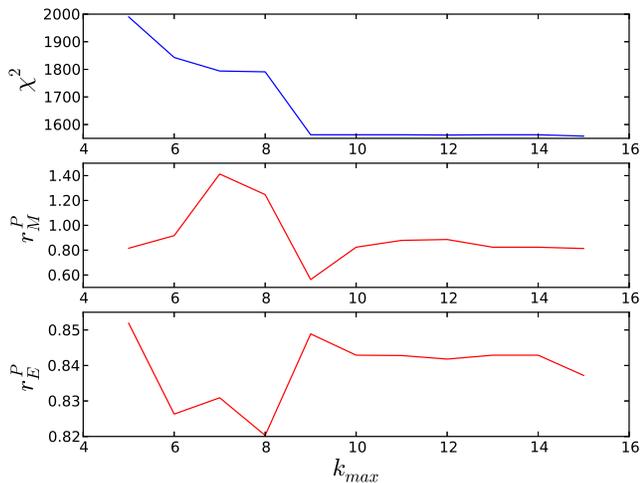}
\caption{Dependence of $\chi^2$, $r_E^P$ and $r_M^P$ on the number 
of terms $k_{\rm max}$ in the expansion in the conformal mapping variable.\label{fig:results}}
\end{figure}
%--------------------------------------------------------------------------------------
To illustrate the comparison to the original analysis \cite{Bernauer:2010wm}, 
we follow the same procedure to choose the number of parameters. This means, 
we increase the number of terms in the expansion Eq.~\eqref{expa} until the 
$\chi^2$ of the fits reach a plateau and the fits stabilize. 
We have performed the calculations in python and checked the results with mathematica. The fits are carried out using the lmfit package with several optimization methods as the Levenberg-Marquardt and simulated annealing algorithms \cite{fitting}.\newline
The extracted 
radii and $\chi^2$-values are shown in Fig.~\ref{fig:results}. 
For $k_{\rm max}=9$, the
absolute $\chi^2$-value of 1563 is reached. This is exactly the value found 
in \cite{Bernauer10} for the best polynomial fit. For $k_{\rm max}=10$, both 
electric and magnetic radii start to stabilize. The proton charge radius is
found to be $r_E^p \simeq 0.84\,$fm, consistent with the muonic hydrogen value
of Ref.~\cite{Pohl:2010zza} and also the one obtained from a dispersion
theoretical analysis of the Mainz and older data, including the ones for the
neutron \cite{Lorenz12}. The level of variation in the magnetic radii is 
larger than in the electric 
case, as we expect, since $G_M$ is suppressed by a factor of the momentum
transfer squared in the reduced cross section, cf. Eq.~(\ref{eq:xs_ros}). 
The proton magnetic radius comes out as 
$r_M^p \simeq (0.85\pm0.04)\,$fm, somewhat larger than the value found by the
A1-collaboration and within large variations compatible with the one obtained 
in Ref.~\cite{Lorenz12}.
The level of precision of the original analysis \cite{Bernauer:2010wm} is reproduced here
only with unconstrained parameters. As an example, we give the parameters for the $k_{\rm max}=10$ fit: $a_1 = -0.9481457, 
a_2=-4.953483, 
a_3=88.55243, 
a_4=-978.0812, 
a_5=6091.365, 
a_6=-22558.25, 
a_7=50007.79, 
a_8=-63978.5, 
a_9=42440.28, 
a_{10}=-10757.15$ for the electric form factor and
$b_1=-2.527861, 
b_2=-12.71964, 
b_3=233.4448, 
b_4=-2025.318, 
b_5=8129.828, 
b_6=-15013.78, 
b_7=4935.569, 
b_8=26389.54, 
b_9=-40617.2, 
b_{10}=18526.76$ for the magnetic form factor. Clearly, such large coefficients generate an unphysical spectral function and thus are far from realistic, just like polynomial or spline fits. However, the crucial point is that constraints on the coefficients as suggested by Hill and Paz increase the $\chi^2$ in our fits. This is also the case in a dispersive framework. Currently, there exist no fits in the literature with an equally good $\chi^2$ to these data that obey all known physics constraints. In order to unambiguously distinguish between purely statistical and theoretically motivated data analyses, here we treat only statistical issues. The consideration of more physically motivated and constrained spectral functions will be covered in a later publication \cite{LDHM} using the statistical reasoning and unphysical fits shown here as a prerequisite. In contrast to here, the constraints from a realistic spectral function and the asymptotic behaviour naively expected from quark counting rules will be 
treated. The possible impact of data at larger $Q^2$ on the radius term is of course an additional source of uncertainty. The main point of this paper is to clearly show that the combined uncertainties of fits without physics input do not allow to distinguish between the `small' and `large' radius.

\medskip
\noindent{\bf 3.}
%-------------------------------------------------------------------
    \begin{figure}[t]
    \centering
    \includegraphics[width=0.48\textwidth]{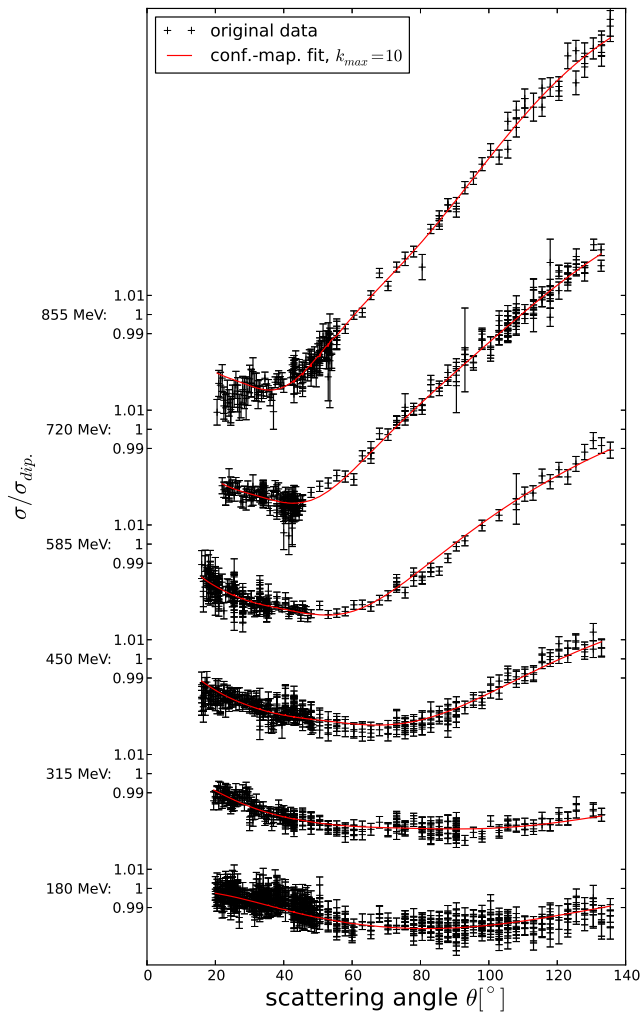}
    \vglue-2mm
    \caption{Cross sections of elastic electron-proton scattering by the 
     A1 collaboration \cite{Bernauer:2010wm}, divided by the cross section of the dipole 
     form factors, $\sigma_{\rm dip}$. All 1422 data points are fitted. The data measured at different
     energies of the incoming electron beam are shown with an offset.
     \label{fig:ccs}}
    \end{figure}
%-------------------------------------------------------------------
The cross sections corresponding to this fit are shown in Fig.~\ref{fig:ccs}
for the conformal mapping function with $k_{\rm max} =10$. 
The six data sets for different energy settings of the incoming electron beam
are  separated by an offset. Note that each of these data sets contains
measurements from three different spectrometers. The different experimental 
settings give rise to a normalization uncertainty between the individual data
sets. One can take this into account by 31 floating normalization parameters, 
as described in the original analysis. According to Ref.~\cite{Bernauer10}, the data contains the normalizations determined by the spline fit. In order to show the underestimation of uncertainties in previous analyses, it is sufficient to keep the normalization parameters fixed. However, the additional normalization uncertainty covers an even larger range of radius values than given here when considered in floating normalizations. To be more precise, we exactly fit to the
same data as was done by the A1-collaboration just using an alternative fit function. 
Other issues like an improved treatment of radiative and
two-photon corrections are not of relevance for this letter but will be taken
up in a later publication \cite{LDHM}, as well as the rigorous inclusion of 
physics constraints. Here, this procedure is necessary for 
a proper comparison to the Mainz analysis.

\medskip
\noindent {\bf 4.}
%-------------------------------------------------------------------
\begin{figure}[tb]
\centering
\includegraphics[width=0.44\textwidth]{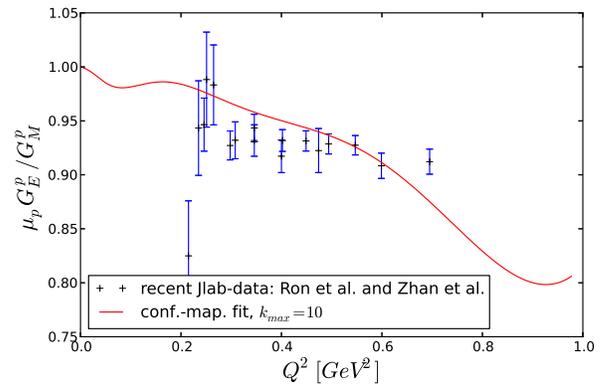} %\hglue-1mm
\caption{Prediction for the form factor ratio from the fit to 
cross sections. The data for the form factor ratio are from polarization 
measurements by Ron et al.~\cite{Ron11} and Zhan et al.~\cite{Zhan11}.
\label{fig:ratio}}
\end{figure}
%-------------------------------------------------------------------
As a further check on our fits, we now consider the form factor ratio that has
been measured precisely using recoil polarization techniques.
The form factor ratio from the illustrative fit with $k_{\rm max} =10$
compares well to the recent measurements at Jefferson Laboratory, see 
Fig.~\ref{fig:ratio}. The displayed ratio is very similar to the one obtained 
in the spline fit in Ref.~\cite{Bernauer10}. The `wiggle' they found below
$Q^2=0.2$~GeV$^2$ from the magnetic form factor is also reproduced. This could
be interpreted as a result from overfitting, which would be consistent with
the fact that the wiggle vanishes when including further physical constraints, 
see e.g. Ref.~\cite{Lorenz12}. We do not want to enter this issue here in more
detail but refer the reader to Ref.~\cite{Meissner:2007tp} for a discussion.
However, under the assumption that all statistical and systematic errors are 
sufficiently under control, a good data description in terms of a low $\chi^2$ 
is required. In this case, one has to consider polynomial, spline and
unconstrained conformal mapping fits on the same footing. In principle, the
latter is to be preferred due to the requirements from analyticity. 
Even if one were to  neglect this fact, one can see from this work that 
a disagreement between the proton charge radius extracted from electron-proton 
scattering data and muonic hydrogen cannot be inferred from 
polynomial or spline fits, as one neglects a sizeable source of uncertainty.

\medskip
\noindent {\bf 5.}
In this Letter, we have reanalyzed the recent elastic electron-proton
scattering  data from Mainz with a fit function that is sufficiently 
flexible to describe the data with a given precision, ie. with the same
precision as achieved by the experimenters using spline and polynomial fit
functions. The results for the proton charge radius $r_E^p$ are in perfect 
agreement with the values obtained via a dispersion relation approach
\cite{Lorenz12} and the recent muonic hydrogen measurements. The 
remaining $r_E^p$-discrepancy is the $\sim 4\sigma$ deviation between the 
average of the spectroscopic measurements in electronic hydrogen and 
those in muonic hydrogen, see e.g. Ref.~\cite{Beyer13}. 
To solve this, further measurements in ordinary
hydrogen are under way~\cite{Beyer13}. The planned muon-proton scattering 
experiment MUSE \cite{MUSE} might also shed further light on these issues.

\bigskip

\begin{acknowledgments}
We thank Jan Bernauer for useful discussions.
We also thank Hans-Werner Hammer for collaboration during the 
initial stage of this project. One of the authors (IL) would like to express a special thanks to the
Mainz Institute for Theoretical Physics (MITP) for its hospitality and
support.
This work is  supported in part by the DFG and the NSFC through
funds provided to the Sino-German CRC 110 ``Symmetries and
the Emergence of Structure in QCD'', and the EU I3HP ``Study of Strongly
Interacting Matter'' under the Seventh Framework Program of the EU.

\end{acknowledgments}

\end{document}